\documentclass[preprint]{aastex631}

\usepackage[utf8]{inputenc}
\usepackage[english]{babel}
\usepackage{natbib}
\usepackage{fp}
\usepackage[capbesideposition=right]{floatrow}
\usepackage{wrapfig}
\usepackage{mathtools}
\usepackage{amsmath}
\usepackage{supertabular,enumitem}
\usepackage{tabularx}
\usepackage{float}
\usepackage{soul}

\submitjournal{ApJ}

\shorttitle{Tracking Non-Radial Outflows}
\shortauthors{Alzate et al.}

\graphicspath{{./}{figures/}}

\begin{document}


\title{Tracking Non-Radial Outflows in Extreme Ultraviolet and White Light Solar Images}

\correspondingauthor{Nathalia Alzate}
\email{nathalia.alzate@nasa.gov}

\author[0000-0001-5207-9628]{Nathalia Alzate}
\affiliation{NASA Goddard Space Flight Center \\
Greenbelt MD 20771, USA}
\affiliation{ADNET Systems, Inc. \\
Greenbelt MD 20771, USA}

\author[0000-0002-6547-5838]{Huw Morgan}
\affiliation{Aberystwyth University \\
Ceredigion, Cymru SY23 3BZ, UK}

\author[0000-0001-6407-7574]{Simone Di Matteo}
\affiliation{NASA Goddard Space Flight Center \\
Greenbelt MD 20771, USA}
\affiliation{The Catholic University of America  \\
Washington, DC, USA}




\begin{abstract}
Understanding the solar corona requires knowledge of its dynamics through its various layers and subsequent connectivity to the heliosphere. This requires understanding the nature of the outflows and the physical transitions through the middle corona ($\sim$1.5-6.0 $R_s$).  While this region is still inaccessible to \textit{in situ} measurements, remote sensing observations are available, but their interpretation can be controversial due to line-of-sight effects and the non-radial motion of outflowing structures close to the Sun ($\leq$3.0 $R_s$). In this work, we describe a method to mitigate these challenges by generating non-radial Height-Time profiles of outflows by using advanced image processing techniques. The North and South boundaries of a large equatorial streamer during the 2008 solar minimum were identified in STEREO/SECCHI solar images, using two different methodologies based on thresholds of brightness and piece-wise polynomial function fitting.  To address line-of-sight issues, we used tomographic reconstruction of the 3D distribution of the coronal electron density based on SECCHI/COR2 images.  Spectral analysis of the time series of the position angle of the streamer boundary revealed its oscillatory nature at some heights at 36-48 hours and 10.5-14.6 hours.  Dividing the distance between the North and South streamer boundaries in equal parts at each height, we obtained non-radial Height-Time paths from which we generated non-radial profiles of corona/solar wind plasma outflow.  We tracked outflows as they moved uninterruptedly from the Sun in EUVI, through COR1 and into COR2.  Finally, we discuss preliminary results of non-radial plane-of-sky velocities for a CME and two small-scale features.

\end{abstract}




\section{Introduction} 
\label{sec:intro}

The magnetic connectivity and physical transitions within the solar corona impact activity throughout the heliosphere.  In this regard, understanding the connection between the low and the high corona is imperative.  Between the two regions, magnetic field configurations transition from predominantly closed to open field.  Studying these regions of the corona is complicated by the fact that neither \textit{in situ} nor single remote sensing instrument observes the corona regularly in its entirety, that is, composite observations from different instruments is currently adopted.  

Despite these shortcomings, many studies have made progress towards a better understanding of the physical properties characteristic of the transition from the low to high corona.  \citet{seaton2021} and \citet{deforest2018}, for example, have described solar wind structures originating in the inner corona and from complex dynamics in the middle corona (defined as the region between heliocentric distances of 1.5 and 6.0 $R_s$ by \citealt{west2022}).  Furthermore, observations by Parker Solar Probe \citep{fox2016} revealed flows and ejecta that interact with the corona's complex magnetic field \citep{howard2019, kasper2019, mccomas2019, bale2019} and are governed by physical transitions between 1.5 and 3.0 $R_s$ \citep{deforest2018, chhiber2019}.  At these heights, the corona is also characterized by the transition from low to high plasma $\beta$ in quiet Sun regions \citep{vourlidas2020}.

Highly non-radial structures in the low corona introduce uncertainties when tracking their outward propagation.  This is typically the case below $\sim$2.5 $R_s$.  However, \citet{boe2020} found that the direction of the coronal magnetic field becomes radial after 3.0 $R_s$.  Using total solar eclipse (TSE) white-light (WL) observations, they inferred the topology of the coronal magnetic field continuously between 1.0 and 6.0 $R_s$.  The scarcity of TSE's, however, motivates other approaches for generating an uninterrupted view of this region of the corona.  Recently, \citet{alzate2021} developed a method, the Bandpass Filtered Frames method (from hereon, the BFF method) to address the high levels of noise in COR1 data (inner coronagraph on STEREO/SECCHI that observes the corona between 1.4 and 4.0 $R_s$).  The method suppresses noise and damps features on timescales based on the chosen filter widths to help reveal outflows propagating from near the Sun out to several solar radii.  However, their approach falls short of capturing non-radial features as they move out of the radial slice field-of-view (FOV) onto which the BFF method is applied.  This paper builds on this by tracking non-radial outflows in extreme ultraviolet (EUV) and WL solar images.  Here, we present our method, which makes use of advanced image processing techniques to identify streamer boundaries in solar images, strengthened by the comparison with boundaries extracted from tomographic brightness reconstruction, from which we generate non-radial Height-Time (Ht-T) profiles of outflows.  Section \ref{sec:data-methods} describes the data used and the details of our method.  Section \ref{sec:results-obs} presents our results followed by our discussion in Section \ref{sec:discussion} and our conclusions in Section \ref{sec:conclusions}.

\section{Data and Methods} 
\label{sec:data-methods}

\subsection{Data and Image Processing}
\label{sec:imgproc}

\begin{figure}[ht!]
        \centering
            \includegraphics[width=\textwidth]{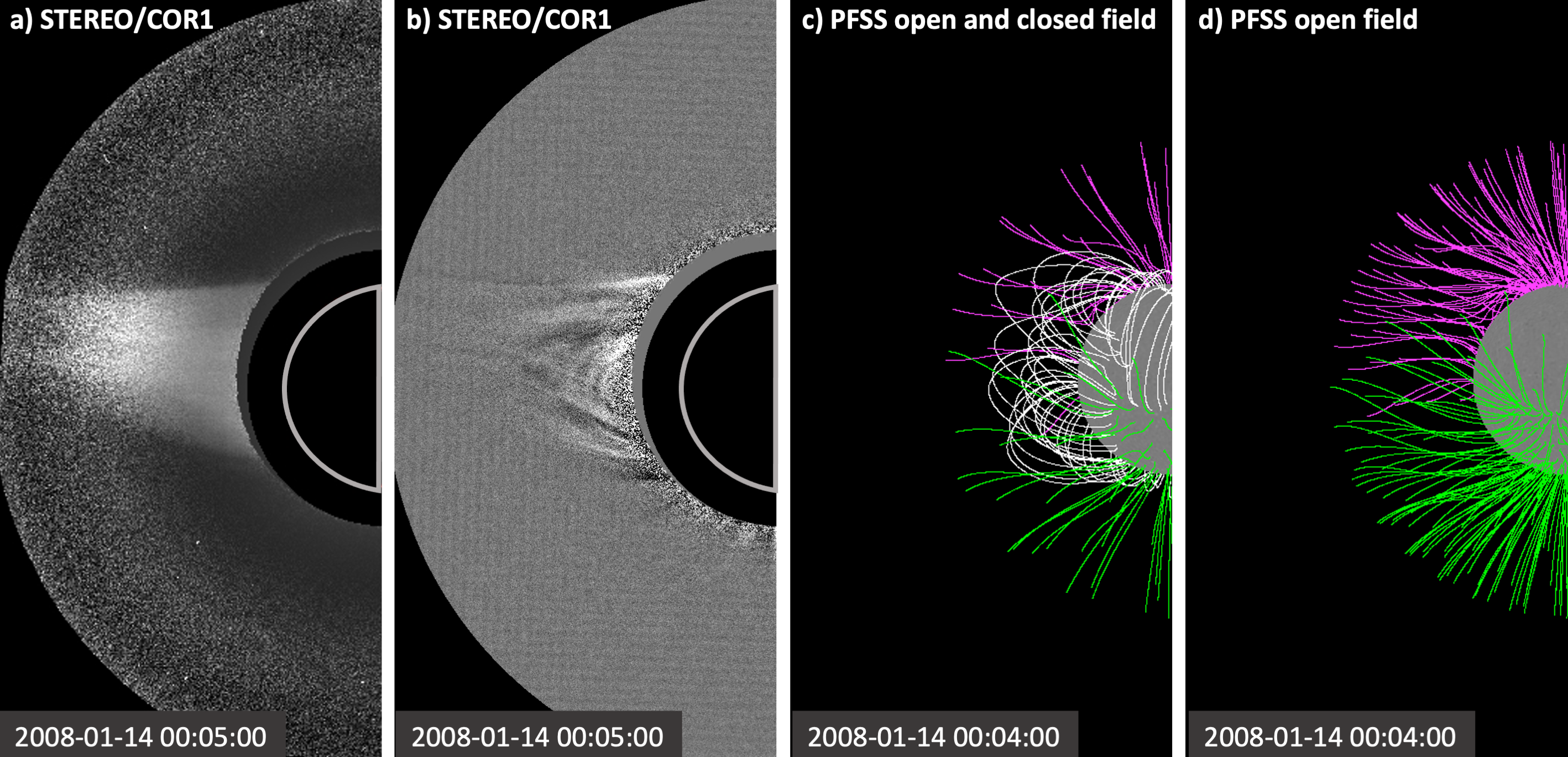} 

            \vspace{0.5cm}
            
            \includegraphics[width=\textwidth]{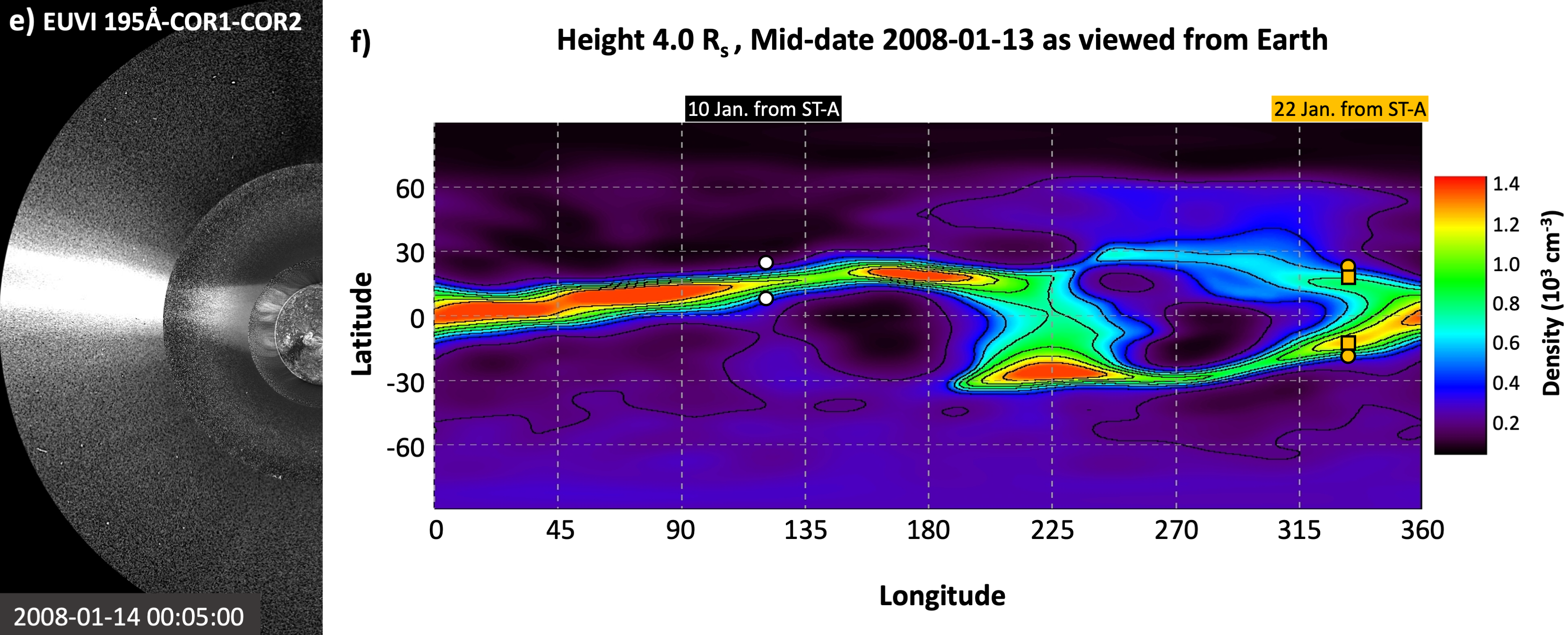}
            \caption{Region of interest showing the streamer belt on 14 January 2008 in a) NRGF-processed COR1 image, b) BFF-processed COR1 image, c) PFSS extrapolation with source surface height at 2.5 $R_s$ showing open and closed field lines, d) same as Panel c) showing only open field lines, e) STEREO/SECCHI composite image and f) tomography reconstruction of the electron density at heliocentric height of 4.0 $R_s$, mapped in Carrington longitude and latitude (note that time goes from right to left). The circles and squares match the ones in Figure \ref{fig:color-profiles}a and \ref{fig:color-profiles}c and correspond to the East limb view of the streamer boundary from STEREO-A on 10 and 22 January.  For the 14 January coronagraph observations in the previous panels, the East limb is at a Carrington longitude of 81$^o$.  Panels a-d) FOV from Sun center out to 4.0 $R_s$.  Panel e) FOV from Sun center out to $\sim$7.0 $R_s$.}
            \label{fig:obs-model-compare}
        \end{figure}

For this study, we chose the same period of low solar coronal activity during 10-23 January 2008 that was analyzed by \citet{alzate2021}.  We used data from the Sun Earth Connection and Heliospheric Investigations (SECCHI; \citealt{howard2008}) suite of instruments onboard the Solar Terrestrial Relations Observatory Ahead and Behind (STEREO-A and -B; \citealt{kaiser2008}) twin spacecraft.  Specifically, we used observations by the Extreme Ultraviolet Imager (EUVI) in the 195 \AA\ channel, the COR1 inner coronagraph \citep{thompson2003} and the COR2 outer coronagraph.  The EUVI instrument observes the Sun and corona out to $\sim$1.7 $R_s$, COR1 observes between 1.4 and 4.0 $R_s$ and COR2 observes between 2.5 and 15 $R_s$.  Together, the three instruments offer an uninterrupted view of the corona as shown in Panel e of Figure \ref{fig:obs-model-compare}.  The data used in this study is from the STEREO-A spacecraft where the streamer analyzed was visible on the East limb as shown.

The methodology we implement here, leverage two known techniques:  (i) the NRGF (Normalizing Radial Graded Filter) method \citep{morgan2006}, which is a simple spatial filter for removing the steep radial gradient of brightness and revealing the electron corona structures; (ii) the BFF (Bandpass Filtered Frames) method \citep{alzate2021}, which is a bandpass filter that operates in the temporal domain to effectively damp high-frequency noise and low-frequency slow-changing structures.  Both the NRGF and BFF methods can be applied to WL and EUV images.  Examples of the application of these two methods are shown in Figure \ref{fig:obs-model-compare} in Panels a (NRGF-processed COR1 image) and b (BFF-processed COR1 image).  Panels c and d compare these observations with the Potential Field Source Surface \citep[PFSS;][]{Schatten1969,Schrijver2003} extrapolation, based on the measurements with SDO/Helioseismic and Magnetic Imager \citep[HMI;][]{Schou2012}, to provide context.  The time series of COR1 images shows the presence of an equatorial streamer belt covering all solar longitudes in the analyzed period.  This is best seen in the tomography map shown in Panel f.  The map shows the Carrington latitude-longitude distribution of the coronal electron density at a heliocentric height of 4 $R_s$ resulting from applying tomography to COR2 A polarized brightness observations made between 2008 January 6-20, using the methods of \citet{morgan2015, morgan2019} and \citet{morgan2020}. For the tomography, input data is required over a period covering at least half a solar rotation (approximately 14 days) in order to ensure complete coverage of the whole corona. This period is centered on the mid-date of 13 January 2008 for the example in the figure. For a tomography reconstruction at a given distance (4 $R_s$ in this example), an annular slice of the polarized brightness images is extracted at that distance. Then, the spherical harmonic-based inversion provides density maps on Carrington longitude-latitude grids of 540 longitude bins and 270 latitude bins. This is a static reconstruction that assumes that the large-scale coronal structure does not change throughout the period of half a solar rotation. The tomography is applied to a set of nine selected heliocentric distances (4.0, 4.4, 5.0, 5.4, 5.9, 6.5, 6.9, 7.5 and 8.0 $R_s$), which will be used in Section \ref{sec:non-rad}. The choice of nine distances is sufficient for our purposes, and this range of distances avoids certain artifacts in the inner FOV of COR2. We restrict the upper distance to 8.0 $R_s$ because the F corona becomes increasingly polarized with distance. We note also that the streamer structure found with the tomography is very similar across this distance range (and to even larger distances) due to the radial expansion of the nascent solar wind.


\subsection{Methods for Generating Non-Radial Profiles} 
\label{sec:non-rad}

\begin{figure}[ht!]
            \centering
               \includegraphics[width=\textwidth]{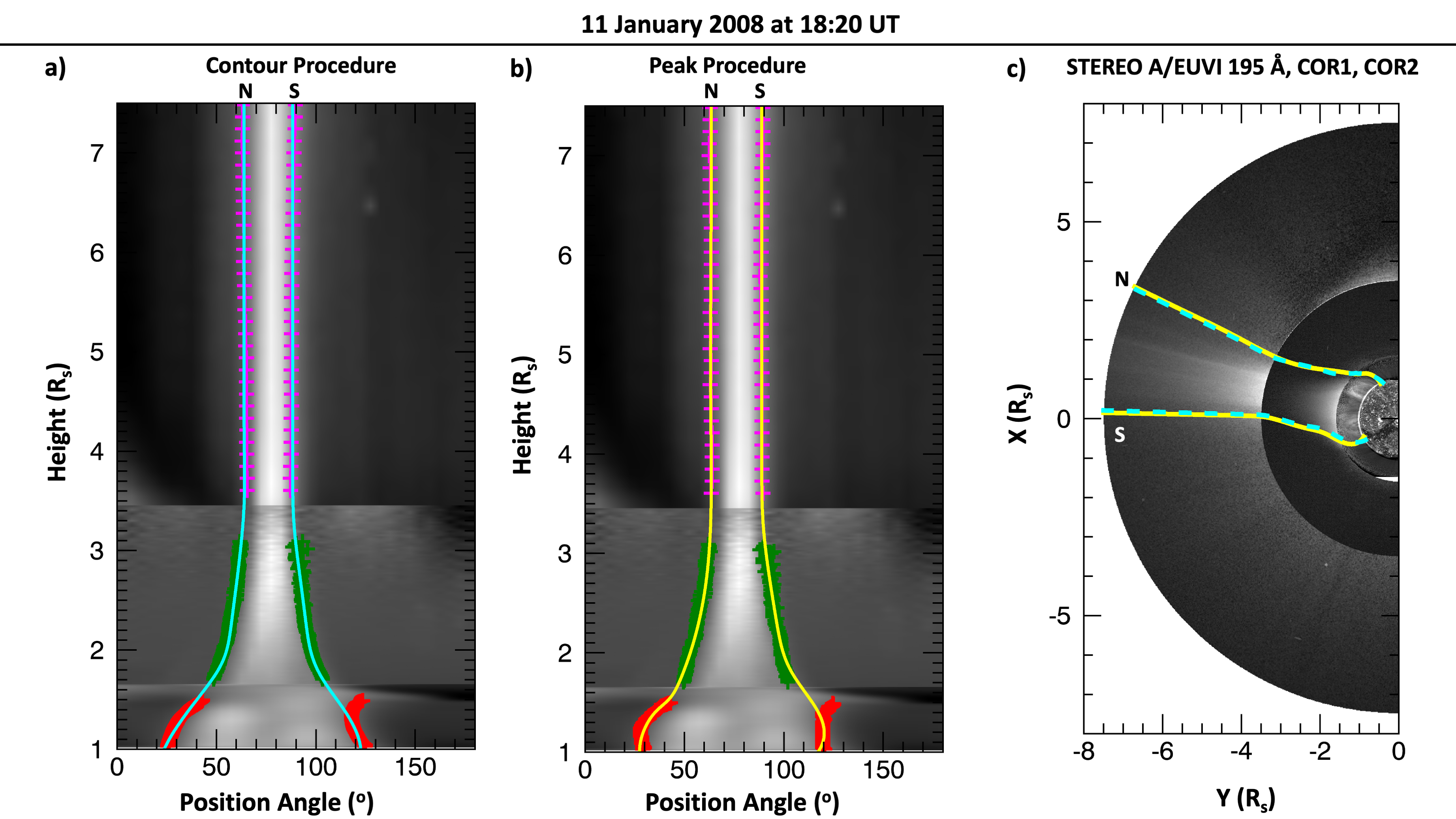}

        \vspace{0.5cm}
        
                \includegraphics[width=\textwidth]{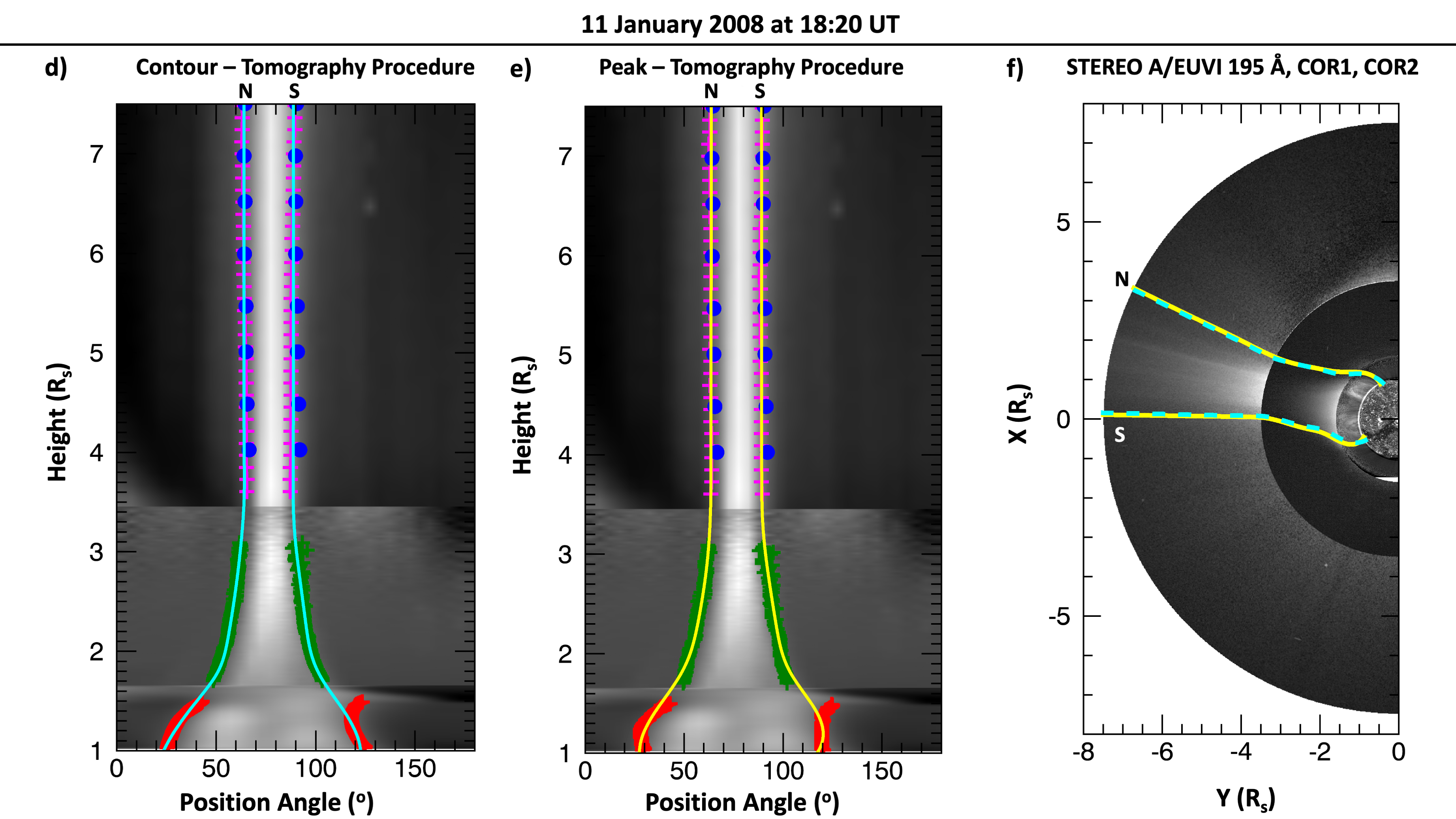}
                \caption{Results for a) \textit{Contour}, b) \textit{Peak}, d) \textit{Contour with Tomography} and e) \textit{Peak with Tomography} methods for 11 January 2008 at 18:20 UT.  Background images are EUVI, COR1 and COR2 composite images in polar coordinates processed with the NRGF method.  c) and f) Final profiles overlay EUVI, COR1 and COR2 composite for corresponding date.  The corresponding streamer boundaries (N for North and S for South) are indicated in all images.  c) and f) COR1 and COR2 are original (calibrated) and unprocessed images; EUVI inlay is processed with the BFF and MGN methods \citep{morgan2014}.}
               \label{fig:boundaries-tomography}
        \end{figure}

We converted the EUVI, COR1 and COR2 observations to polar coordinate images rebinned in uniform steps of position angle (\textit{pa}; positive counterclockwise from north) and height (\textit{h}). We divided the FOV of each instrument into 72 \textit{pa} bins and 100 \textit{h} bins. The resulting images are averaged over four hours, specifically 48 images for EUVI and COR1 at 5 minute cadence, and 12 images for COR2 at 20 minute cadence. The choice of four hour intervals is to enable enough signal to identify the streamer boundaries without sacrificing the details of their evolution.  We obtain a robust representation of the observed large equatorial streamer boundary every four hours from 10 January 2008 at 00:05 UT to 22 January 2008 at 00:05 UT.  We first applied the NRGF method to reveal the streamer structure in the observations (background images shown in Panels a-b and d-e of Figure \ref{fig:boundaries-tomography}).  For EUVI observations, an additional Gaussian smoothing via the routine {\fontfamily{cmtt}\selectfont GAUSS\_SMOOTH} (within the \textit{Interactive Data Language} (IDL) based system) with standard deviation set to 2.0 was used to better reveal the large-scale structure of the streamer by reducing variations due to smaller scale gradients.  

Capitalizing on the uninterrupted 1–15 $R_s$ radial FOV provided by the SECCHI suite of instruments, we developed two procedures for the identification of the non-radial streamer boundaries.  For the first method (hereafter referred to as \textit{contour} method), we imposed a threshold on the brightness of the entire image (independently for EUVI, COR1 and COR2) defined as a certain percentile of the brightness values.  The border of the regions with values above the threshold define the streamer boundaries.  This method provides closed contours that we separate into two distinct profiles (north and south profiles) considering only the points within a maximum and minimum height for each instrument.  These points are shown in Figure \ref{fig:boundaries-tomography}a with crosses (red crosses in EUVI, green in COR1 and magenta in COR2).  For the second method (hereafter referred to as \textit{peak} method), we extracted, at each height, the brightness profile as a function of the position angle and searched for local peaks with respect to a threshold defined as a certain percentile of the brightness values at that height.  For a finer identification, the brightness profile is linearly interpolated to increase the number of data points by a factor of 10 (the choice of this factor is arbitrary and does not affect the results).  Repeating the analysis within the same height range as for the \textit{contour} method, and for each instrument, yielded a series of points representing the preliminary streamer non-radial profiles.  These are shown in Figure 2b with crosses (red crosses in EUVI, green in COR1 and magenta in COR2).  For both methods, we chose the percentile threshold empirically, lowering the value until each methodology identified the full streamer instead of isolated brighter portions of the streamer itself.  We chose the 72\% and 80\% percentile levels for coronagraph observations, respectively for COR1 and COR2, and the 38\% level for EUVI observations (AIA 195 \AA).  We limited the analysis to 1.03-1.5 $R_s$ for EUVI, 1.7-3.1 $R_s$ for COR1 and 3.6-7.5 $R_s$ for COR2, which removes the area of the image most affected by noise near the edges of the instruments' FOV.

To obtain a continuous profile, we proceeded with a piece-wise fitting methodology, independently for the two boundaries of the streamer.  We used a third order polynomial for the boundary between the EUVI FOV and the lower 20 points of the COR1 FOV for the \textit{contour} method and lower 40 points for the \textit{peak} method (eq. 1). We used another third order polynomial for the remaining points of COR1 and the first 10 points of COR2 (eq. 2). For the remaining points of COR2, we simply considered the average value of the boundary position since, at this step, the profile has reached the radial portion of the corona (eq. 3),

    \begin{align}
        pa(h)_{EUVI,COR1} = a_0 h^3 + b_0 h^2 +c_0 h + d_0\\
        pa(h)_{COR1,COR2} = a_1 h^3 + b_1 h^2 +c_1 h + d_1\\
        pa(h)_{COR2} = d_2 
    \end{align}

\noindent where the streamer profile is expressed in terms of position angle (\textit{pa}) versus height (\textit{h}) and $a_i$, $b_i$, $c_i$ and $d_i$ are the free parameters.  A unique function for the non-radial profile was then obtained connecting the piece-wise fit with sigmoid functions, $\alpha(h)$ and $\beta(h)$, as follows:

    \begin{align}
        \alpha(h) = \frac{1}{1+e^{(h-h_{\alpha})/\lambda_{\alpha}}}\\
        \beta(h) = \frac{1}{1+e^{(h-h_{\beta})/\lambda_{\beta}}}\\
        pa(h)_{non-radial} = (1-\alpha)\,pa(h)_{EUVI,COR1} + \alpha[(1-\beta)\,pa(h)_{COR1,COR2} + \beta \,pa(h)_{COR2}] 
    \end{align}

Note that in this last step, only the sigmoid functions have free parameters ($h_{\alpha,\beta}$ and $\lambda_{\alpha,\beta}$) while the coefficient $a_i$, $b_i$, $c_i$ and $d_i$ are kept fixed from the previous step. Following this procedure, we generated non-radial streamer profiles every four hours. The fitting method was applied to each of the two methods (\textit{contour} and \textit{peak}).  The non-radial profile generated is shown in blue in Figure \ref{fig:boundaries-tomography}a for the \textit{contour} method and in yellow in Figure \ref{fig:boundaries-tomography}b for the \textit{peak} method.  The same profiles are shown in Figure \ref{fig:boundaries-tomography}c overlaying the image composite (EUVI, COR1 and COR2) with dotted blue and yellow lines.  Corresponding profiles are indicated with ``N" for North and ``S" for South.  We refer to the two boundaries as north (N) and south (S) profiles since the position angle is defined as positive counterclockwise from north.  The position angle of the non-radial profiles at a cadence of four hours can be linearly interpolated in time at each height to obtain profiles at the cadence of the instrument of interest (i.e., EUVI, COR1, COR2).

Using latitudinal profiles from tomography reconstructions of brightness at fixed heights (proportional to the density values in Figure \ref{fig:obs-model-compare}f), we extracted a brightness profile as a function of position angle.  We did this for a given Carrington longitude corresponding to the FOV plane of STEREO-A at a given time.  Then, we selected the peak in brightness, corresponding to the streamer, using the same selection criteria as that of the \textit{peak} method described in Section \ref{sec:non-rad}.  Available tomography maps cover nine heights in the COR2 FOV and hence provide nine additional points per boundary, shown as blue filled circles in Panels d–e (the points at 8.0 $R_s$ are not shown in the figures).  These additional points were included with the ones obtained from the \textit{contour} and \textit{peak} methods to provide more robust results.  Namely, we used the lower four points (4.0 to 5.4 $R_s$) for the polynomial fit between COR1 and COR2, and the other five points (5.9 to 8.0 $R_s$) for the average value in the COR2 FOV. Indeed, information from tomography maps partially accounts for line-of-sight uncertainties.  The newly generated profiles are shown in Panels d–f in the same format as Panels a–c (blue and yellow lines).  

The methodology described in this section can be summarized in the following steps:

\begin{enumerate}[itemsep=0mm]
    \item Convert images to polar coordinates rebinned (\textit{pa}=72, \textit{h}=100).
    \item Average images over four hours.
    \item Streamer identification:
    \begin{enumerate}[itemsep=0mm]
        \item \textit{Contour}---impose a threshold on the brightness of the entire image.
        \item \textit{Peak}---extract the brightness profile as a function of \textit{pa} at fixed heights and search for local peaks.
    \end{enumerate}
    \item Apply a piece-wise polynomial fit, independently for the two streamer boundaries (between EUVI and COR1; between COR1 and COR2; on COR2 only).
    \item Connect the piece-wise polynomial fit with sigmoid functions at each boundary independently.
    \item Generate non-radial streamer profiles every four hours.  Linearly interpolate in time at each height for higher cadence.
    \item Optional step---Using tomography, extract streamer boundaries at a fixed height and time from a brightness profile as a function of \textit{pa}.  
    \item Construct a datacube following the steps described in Section \ref{sec:datacube}.
    \item Generate non-radial Ht-T plots from the datacube in Step 9 and following the steps in \citet{alzate2021}.
\end{enumerate}


\subsection{Method for Generating Datacubes} 
\label{sec:datacube}
We rely on the non-radial streamer boundaries to define \textit{non-radial path} bins.  Specifically, at each height, the angular distance between the North and the South boundaries is divided into $n_{path}$ bins of uniform angular size.  This number of bins is the same at all heights, so that their angular size decreases with height in the case of the equatorial streamer analyzed here.  For this work, we defined seven paths obtaining an angular size in the COR2 FOV ranging between $\sim$3$^\circ$ and $\sim$7$^\circ$ (see Table \ref{tab:pa_ranges}).  These values are comparable to the angular width of the \textit{radial} slices used by \citet{alzate2021}, that is 5$^\circ$. This enabled the comparison between results from the \textit{radial} datacube in \citet{alzate2021} and the \textit{non-radial} datacube in this work.  Each path was subsequently separated into bins of uniform $n_h$ steps in height.  These bins were finally used to build a datacube with dimensions $[n_t,n_h,n_{path}]$ storing spatial average image intensities from EUVI, COR1 and COR2 observations (original, calibrated, unprocessed images) with a cadence of 5 min, 5 min and 20 min, respectively. Time data gaps in the datacube, due to gaps in observation, amounted to 5.1\% in EUVI (primarily on 20–21 January), 2.8\% in COR1, and 0.9\% in COR2 and were filled by linear interpolation.
\section{Results/Observations}
\label{sec:results-obs}


    \begin{figure}[b!]
        \centering
            \includegraphics[width=\textwidth]{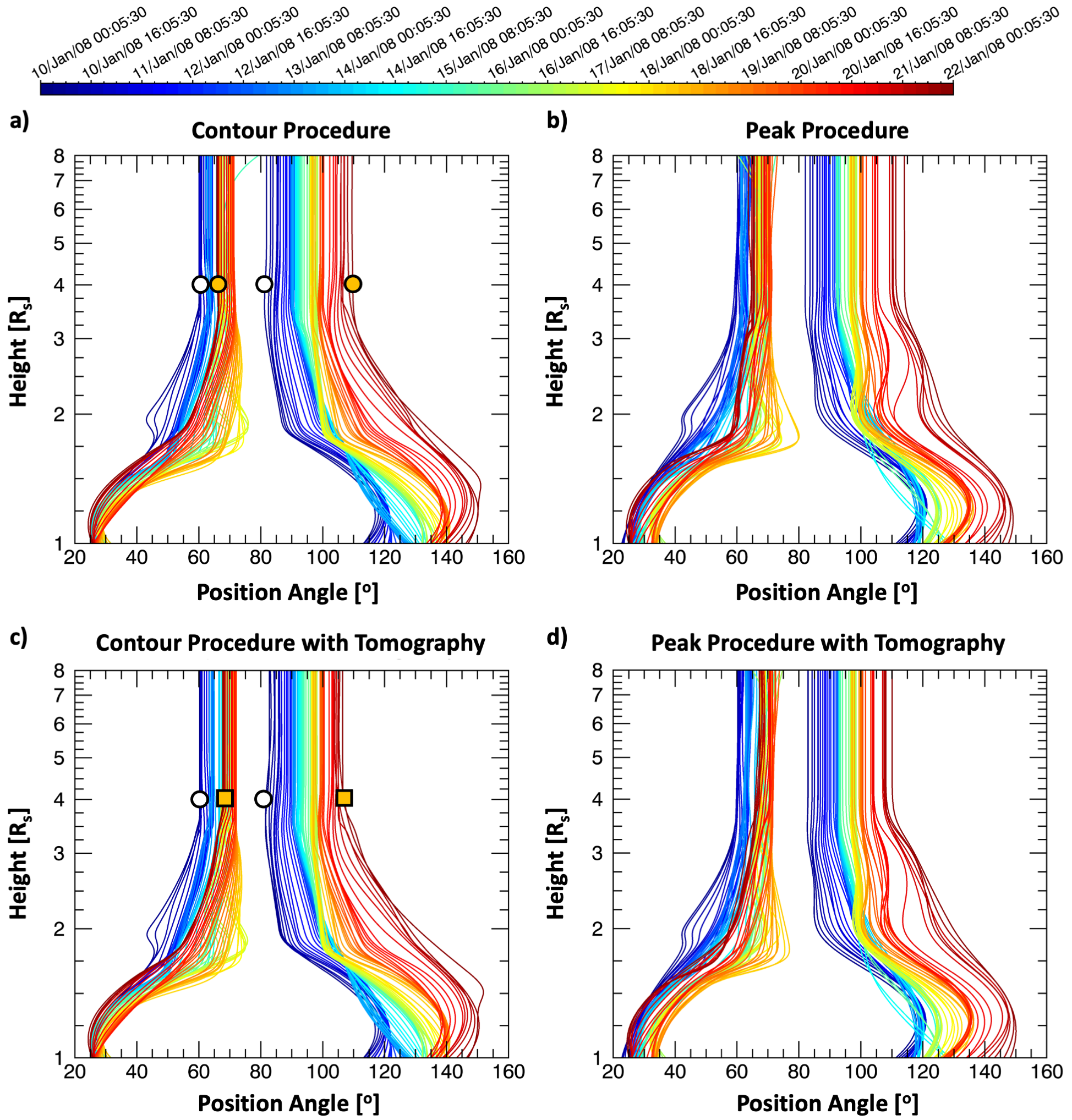}
            \caption{Time evolution of streamer boundary profiles for each method shown for EUVI, COR1 and COR2 up to 8.0${R_s}$.  The different colors indicate the streamer profile at different times.  The white and yellow circles and squares correspond to those indicated in Figure \ref{fig:obs-model-compare}f.}
            \label{fig:color-profiles}
        \end{figure}

\subsection{Time Evolution of Streamer Boundary Profiles}
\label{sec:evolution}

Our methodology enabled the study of the evolution of the streamer boundaries.  To do this, we used the profiles generated every four hours for the entire 12-day time interval in our analysis, which are shown in Figure \ref{fig:color-profiles}.  The different colors indicate the streamer profile position at different observation times.  Panels a and b show the evolution of the boundaries obtained via the \textit{contour} and \textit{peak} methods, respectively, while Panels c and d show the same but with tomography information included.  Note that the streamer boundaries were identified in the plane-of-sky, hence variations in the profiles with time reflect both the changing projection effect due to rotation, as well as actual time evolution of the streamer.

\textbf{\textit{Contour vs. Peak}}---Near Sun surface, regardless of the methodology, the North boundary is confined to within 20$^o$–35$^o$ in position angle.  The South boundary gradually moves from $\sim$110$^o$ towards higher position angles with time indicating the streamer expands with decreasing Carrington longitude.  The position angle of the boundaries with changing height exhibits the typical helmet streamer shape.  The profile widths shrink significantly within the FOV of COR1, but the non-radial nature of the profiles is preserved until $\sim$3.0 $R_s$, which is in agreement with the findings by \citet{boe2020}.  Beyond 3.0 $R_s$, the profile is observed to be radial.  Note that differences between the boundary locations, as determined by the different methods, are more noticeable within the FOV of COR1, possibly due to the higher level of noise in this instrument.  While most of the profiles obtained from the \textit{contour} method are smooth, the ones obtained from the \textit{peak} method show a large number of undulated profiles.  In the COR2 FOV, the North boundary evolves differently from the South boundary.  The South boundary progressively moves at larger position angles, from about $\sim$80$^o$ to $\sim$110$^o$, following the overall expansion of the streamer for decreasing Carrington longitude, also visible in Figure \ref{fig:obs-model-compare}f.  In particular, the tomography results at a heliocentric height of 4.0 $R_s$ show a high density region progressively wider in latitude from January 10 to January 22.  This is indicated by white/orange circles at the beginning/end of the time interval corresponding to the circles shown in Figure \ref{fig:color-profiles}a.

For the North boundary, some profiles overlap.  This is because the boundary oscillates within a smaller range of position angles, from $\sim$55$^o$ to 70$^o$.  This is observed (in the COR2 FOV) in the profile extracted by both methods and is even more evident in the COR1 FOV where the red profile is confined between blue and orange profiles.  Interestingly, the North region of the streamer is where a coronal mass ejection (CME) crossed the COR1 FOV during 16 January 2008, suggesting a possible relation between the streamer boundary oscillations and the CME propagation.  This day also corresponds to the more undulated North profiles from the \textit{contour} procedure (light green profiles in Panel a).  

\textbf{\textit{Contour with Tomography} vs. \textit{Peak with Tomography}}---Panels c and d in Figure \ref{fig:color-profiles} show very similar results for the profiles obtained by the two methods when data points from tomography maps are included.  Overall, the set of profiles appear smoother than those obtained without tomography, especially in the COR2 FOV where the position angle distribution of the profiles is more compact and closer to the higher density region of the streamer as identified by the tomography reconstruction.  This can be seen by comparing the location of the streamer profiles at 4.0 $R_s$ (orange squares in Figure \ref{fig:color-profiles}c) with the ones obtained without tomography (orange circles in Figure \ref{fig:color-profiles}a) also shown in the tomography 3D density reconstruction (Figure \ref{fig:obs-model-compare}f).  The additional constraint from the tomography improves the reliability of the fit in the COR1 FOV resulting in less undulated profiles.  Finally, there are no major differences among the profiles in the EUVI FOV.  

\begin{figure}[t!]
        \centering
            \includegraphics[width=\textwidth]{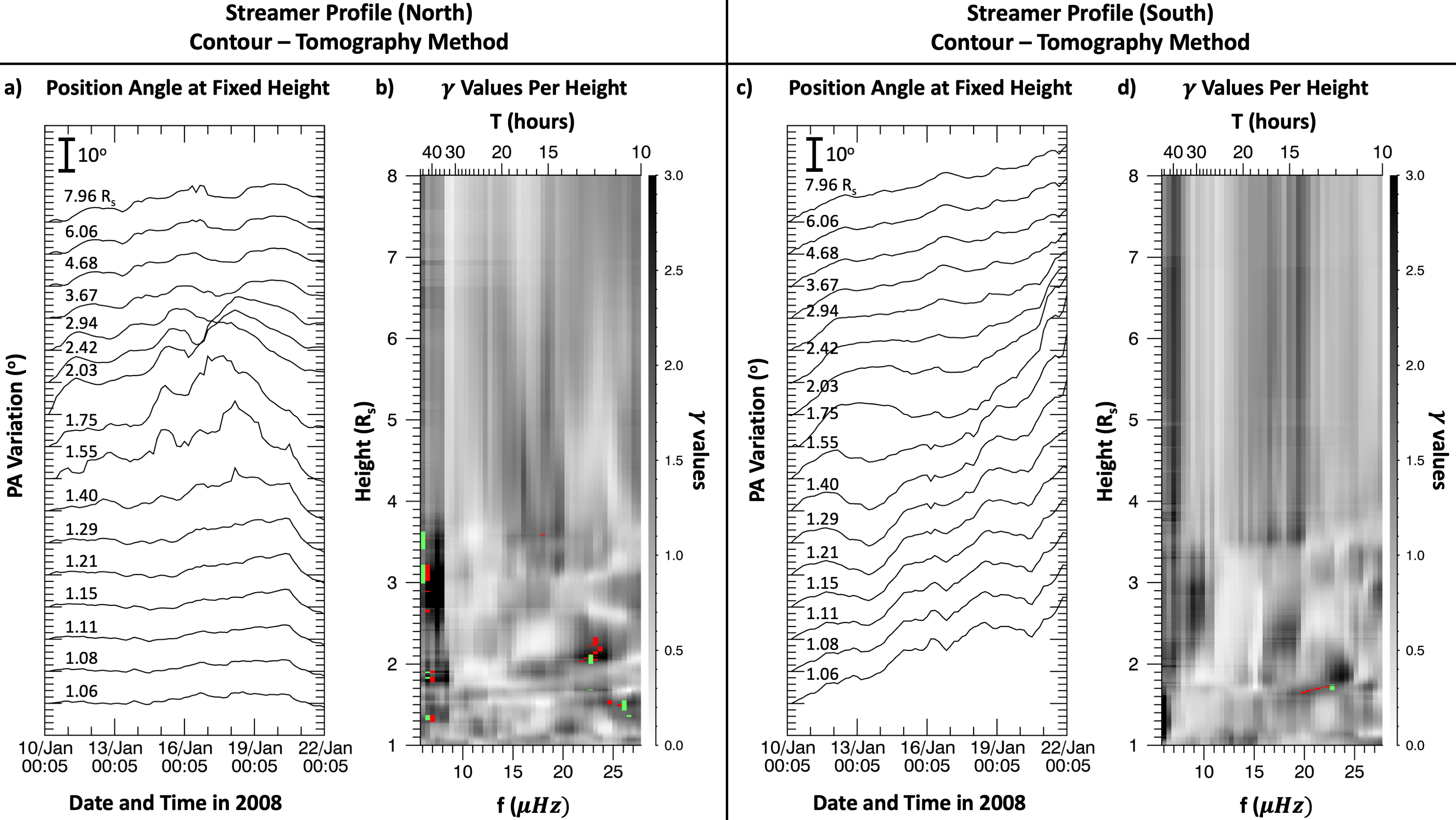}
            \caption{Position angle values as a function of time for the \textit{contour with tomography} method.  a) and c) Profile variations for the North and South boundaries at selected heigths.  b) and d) Results from the spectral analysis.  Color scales represent the ratio between the PSD and the estimated background for \textit{pa} time series at all heights.  The red dots identify the time and the center frequency of the power enhancements above 90\% confidence threshold ($\gamma$ test).  The green dots mark the portions simultaneously passing the F test.}
            \label{fig:evolution-profiles}
        \end{figure}

\textbf{\textit{Position Angle Profiles}}---To better characterize the oscillations of the streamer boundaries and the role of the CME, we extracted the position angle values as a function of time at fixed heights spanning the FOV of the three instruments.  Figure \ref{fig:evolution-profiles} shows the profile variations in the North (Panel a) and South (Panel c) boundaries for the results from the \textit{contour with tomography} method.  In the North profiles, there are no major position angle variations below 1.4 $R_s$, but oscillations occur above this height; in the South profiles, upward trends in position angle variations occur at all heights.  Additionally, it is noticeable that for the North profiles, the oscillations are stronger at heights between 1.4 and 3.0 $R_s$.  Also, two strong position angle variations observed in the profile at 1.75 $R_s$ can be seen in a broader form at later times at both smaller and larger heights suggesting a propagation of a disturbance generated at $\sim$1.8 $R_s$.  Note that these peaks are observed around the time of the CME launch, further suggesting that the CME plays a role in the deformation of the streamer boundary.  In addition, during the same day, we observed a smaller peak in the variations of the South profile position angle that is progressively broader and occurs later in time with increasing height.  

We performed a spectral analysis of the position angle variations using the procedure by \cite{dimatteo2021}.  Briefly, after zero-padding the time series to reach two times the original length, we estimated the power spectral density (PSD) via the adaptive multitaper method \citep[MTM;][]{thomson1982} with time-halfbandwidth product $NW=3$ and number of tapers $K=5$.  Then, via a maximum likelihood criterion, we estimated a continuous PSD background fitting a bending-power-law function to the PSD smoothed with a running geometric mean (bin+BPL combination in \citealt{dimatteo2021}).  The ratio between the PSD and the estimated background constitutes the $\gamma$ values. We impose a 90\% confidence threshold to identify the frequency of significant PSD enhancement signature of periodic fluctuations in the time series ($\gamma$ test).  The results are combined with the harmonic F test, an additional independent statistical test provided by the multitaper methodology, thus strengthening the reliability of our results.

We applied this procedure to the position angle time series to estimate the $\gamma$ values for each height in the EUVI, COR1 and COR2 FOV.  This is represented with color scale in Panel b and d of Figure \ref{fig:evolution-profiles}.  We considered the entire time interval available (73 data points) that, with a sampling time of four hours, determined a Nyquist frequency of $\approx$34.7 $\mu$Hz and a Rayleigh frequency of $\approx$0.95 $\mu$Hz. Based on these values and the choice of parameters for the analysis, the reliable frequency range unaffected by border effects is 5.80–27.8 $\mu$Hz, that is, periods between 10 and 48 hours \citep{dimatteo2021}, hence the limited frequency/time range in Panels b and d.  The red and green dots indicate the portion of the spectra above the imposed 90\% confidence level for the $\gamma$ and F-test, respectively.  In the North profile, we identified periodic fluctuations of the streamer boundary between 1.4–1.6 $R_s$ at 24.5–26.5 $\mu$Hz (10.5–11.3 hours) and between 2.0–2.4 $R_s$ at 22.0–23.0 $\mu$Hz (12.1–12.6 hours). In the South profile, the $\gamma$ test revealed fluctuations at frequencies from 19 $\mu$Hz (14.6 hours) at 1.7 $R_s$ to 23 $\mu$Hz (12.1 hours) at 1.8 $R_s$.  The combination with the F-test constrained the frequency values to a narrower range, namely 22.0-23.0 $\mu$Hz (12.1–12.6 hours).  An additional periodicity at longer time scales is identified only in the North profile at about 6.0–8.0 $\mu$Hz (34.7–46.3 hours). However, darker regions in the $\gamma$ values can also be recognized in the results for the South profiles at almost all heights.  Some indication of such $\approx$1.5–2 day fluctuations is visible in the position angle time series (Panel a and c).


\subsection{Non-Radial Height Time Profiles for Each Method}
\label{sec:htt-plots}

\begin{figure}[ht!]
        \centering
            \includegraphics[width=\textwidth]{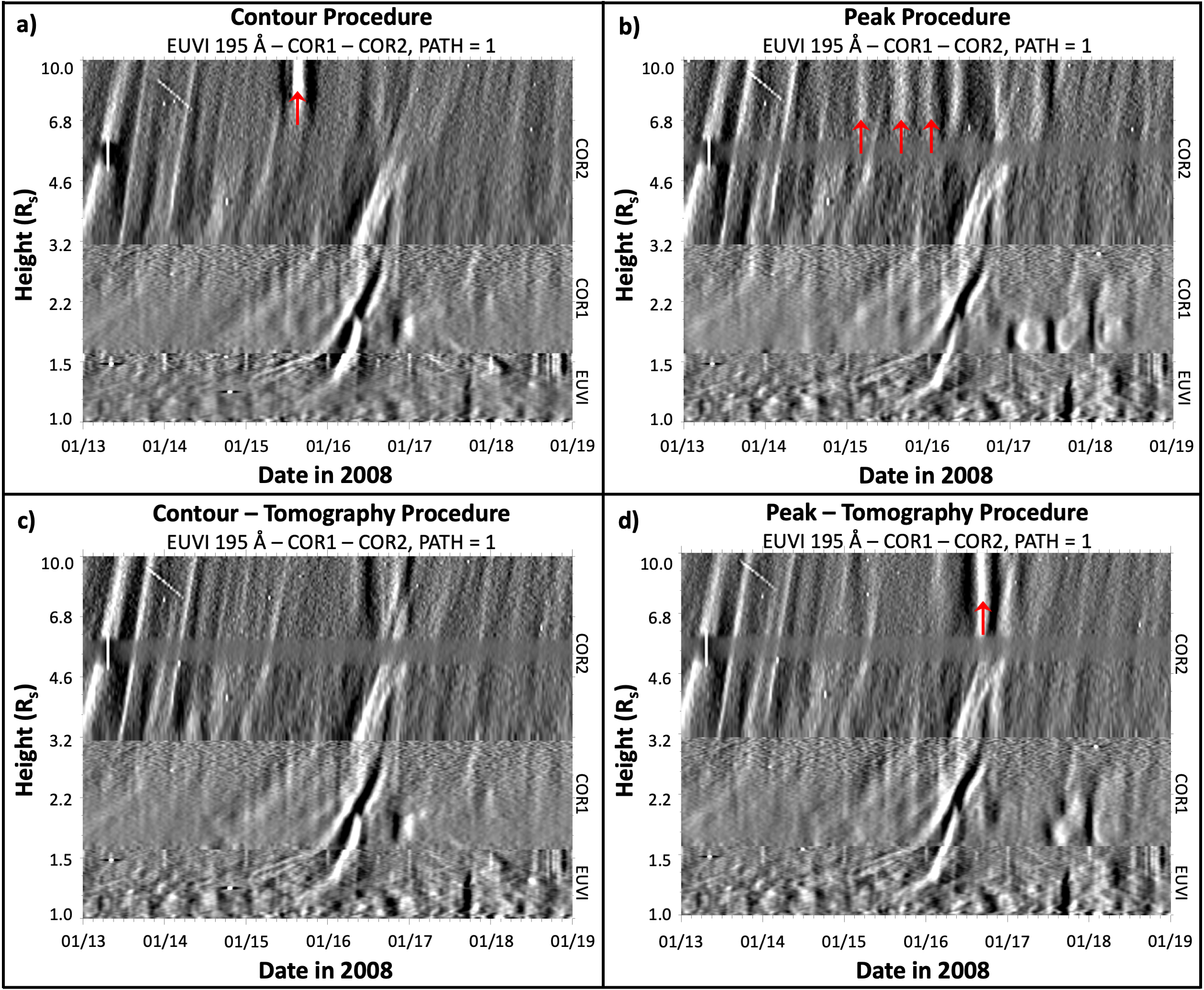}
             \caption{EUVI, COR1 and COR2 Ht-T profiles of non-radial Path=1 obtained from the \textit{contour} and \textit{peak} methods with and without tomography.  Shown is a shorter time interval from 13 to 19 January 2008 (centered around the 16 January CME).}
            \label{fig:all-meth-htt-plots}
        \end{figure}

Using the procedure described in Section 2.3, we generated datacubes of time, height bins and path bins, starting from the original, calibrated, unprocessed images.  We then applied the BFF method (as described in \citealt{alzate2021}) with wide and narrow filters obtained via the IDL {\fontfamily{cmtt}\selectfont GAUSS\_FUNCTION}, respectively, with standard deviations equal to 20 and 5 for both EUVI and COR1.  These choices damp time scales respectively greater than 10.56 hours and lower than 2.5 hours.  For COR2, the choice of filters was of standard deviation 4 (wide) and 0.4 (narrow), which damped time scales respectively greater than 9.5 hours and lower than 3.5 hours.  Using each datacube generated from the \textit{contour} and \textit{peak} procedures, with and without tomography, we generated Ht-T plots for each non-radial path.  As an example, Figure \ref{fig:all-meth-htt-plots} shows the four plots corresponding to Path 1, which corresponds to the second path starting from the North boundary of the analyzed streamer (Path 0 is the northernmost path).  Figure \ref{fig:all-meth-htt-plots} shows the results for a time interval shorter than our 12 days of interest, namely 13-19 January 2008 (centered around the time the CME launched).  Following the evolution of the streamer, the non-radial path covered progressively different position angles.  Table \ref{tab:pa_ranges} summarizes, for each methodology, the position angle range of Path 1 at the beginning and end of the time interval analyzed and at heliocentric heights of $\sim$1.0 $R_s$ and $\sim$8.0 $R_s$.

\begin{table}[h!]
    \centering
    \begin{tabular}{c||c|c||c|c||c|c||c|c}
             & \multicolumn{2}{c}{\textit{contour}} \vline\,\vline & \multicolumn{2}{c}{\textit{peak}} \vline\,\vline & \multicolumn{2}{c}{\textit{contour+tomo.}} \vline\,\vline & \multicolumn{2}{c}{\textit{peak+tomo.}} \\
    \hline
             & $\sim$1.0 $R_s$ & $\sim$8.0 $R_s$ & $\sim$1.0 $R_s$ & $\sim$8.0 $R_s$ & $\sim$1.0 $R_s$ & $\sim$8.0 $R_s$ & $\sim$1.0 $R_s$ & $\sim$8.0 $R_s$ \\
    \hline
    \hline
     Jan. 10 & 38.5$^\circ$–50.9$^\circ$ & 63.2$^\circ$–66.3$^\circ$ & 35.6$^\circ$–48.2$^\circ$ & 63.2$^\circ$–66.4$^\circ$ & 38.5$^\circ$–50.9$^\circ$ & 63.5$^\circ$–66.7$^\circ$ & 35.6$^\circ$–48.2$^\circ$ & 63.5$^\circ$–66.7$^\circ$ \\
     
     Jan. 22 & 42.9$^\circ$–60.3$^\circ$ & 72.1$^\circ$–78.3$^\circ$ & 42.5$^\circ$–60.1$^\circ$ & 71.1$^\circ$–78.2$^\circ$ & 42.9$^\circ$–60.3$^\circ$ & 73.3$^\circ$–78.8$^\circ$ & 42.7$^\circ$–60.3$^\circ$ & 73.3$^\circ$–79.4$^\circ$ \\
     \hline
    \end{tabular}
    \caption{Position angle range of Path 1 for each method at specific heliocentric height and time.}
    \label{tab:pa_ranges}
\end{table}

This comparison helps determine the performance of each procedure and identify potential artifacts introduced by our methodology.  For the latter, we recognize vertical features alternating sharp variations of brightness, from very faint to very bright, that do not match when comparing all four panels.  These are easily recognized above $\sim$7.0 $R_s$ (red arrows) in Panel a (15 January at approximately 15:00 UT), Panel b (from 15-17 January) and Panel d (16 January at approximately 15:00 UT).  Similar features occurred between 1.5 and 2.0 $R_s$ in all panels after 16 January at approximately 15:00 UT, with the plot from the \textit{peak} procedure having the most marked artifacts.  Comparison with Figure \ref{fig:color-profiles} reveals that these artifacts are likely related to sudden strong variations in the streamer profiles.  Note, for example, the artifact in Panel a of Figure \ref{fig:all-meth-htt-plots} (15 January at approximately 15:00 UT around 7.0 $R_s$).  This corresponds to the curved green line in Figure \ref{fig:color-profiles}a above 7.0 $R_s$ at around the same time.  Similar characteristics are related to the other artifacts mentioned.  Less marked variations in the streamer profiles can introduce changes in the contrast of subsequent non-radial paths.  We can see this effect in all panels in Figure \ref{fig:all-meth-htt-plots} in the COR1 FOV between 15 and 16 January.  However, underlying slopes of bright features indicate coronal plasma outflows are preserved, hence confirmed, by all the procedures.  Given the lower number of artifacts and more robust (less undulated) profiles, we identify the \textit{contour with tomography} method as the best option for generating non-radial profiles.


\subsection{Comparison Between Non-Radial and Radial Height-Time Plots}
\label{sec:compare-htt}

\begin{figure}[ht!]
        \centering
            \includegraphics[width=0.52\textwidth]
            {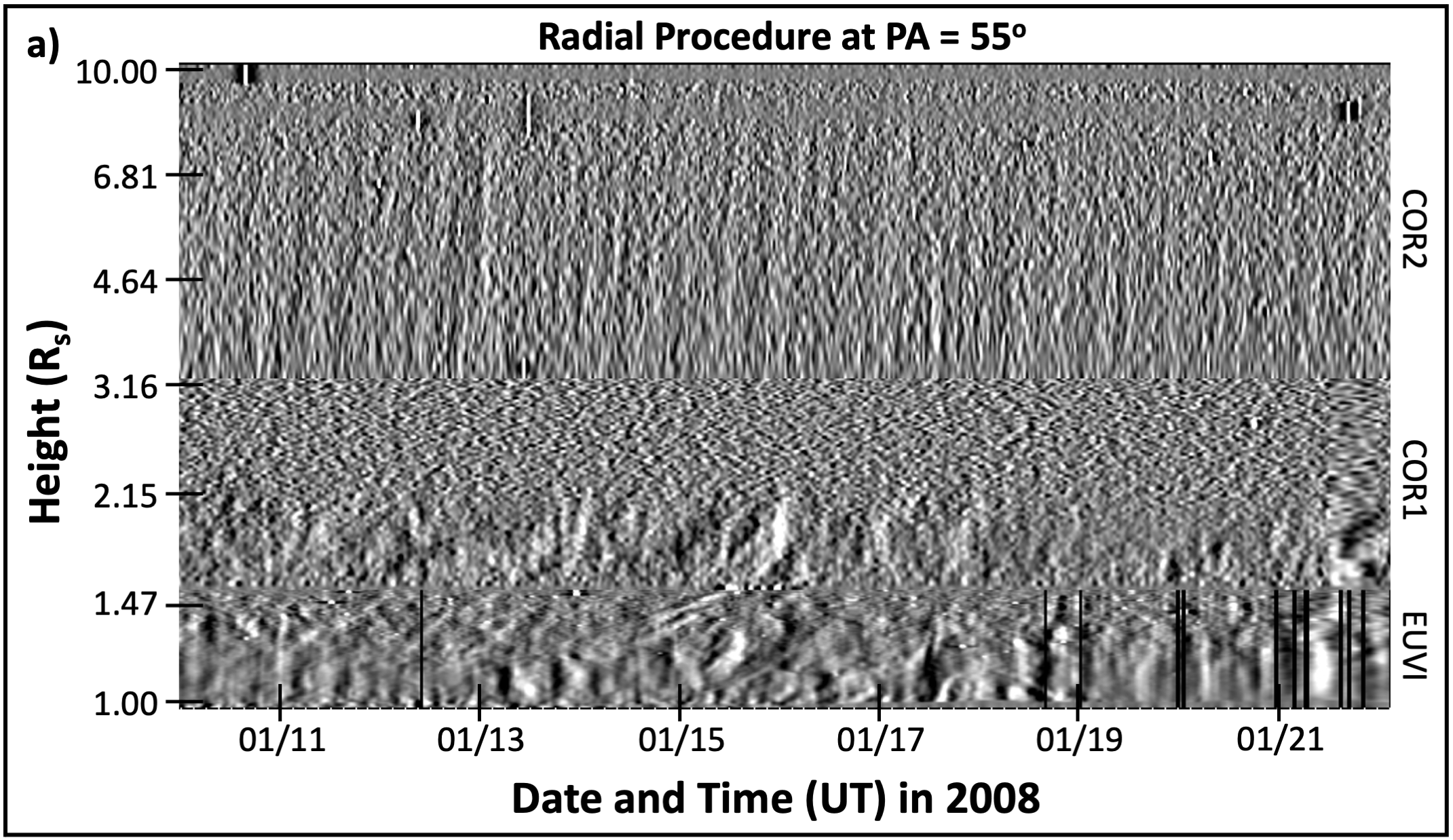}
            \includegraphics[width=0.52\textwidth]{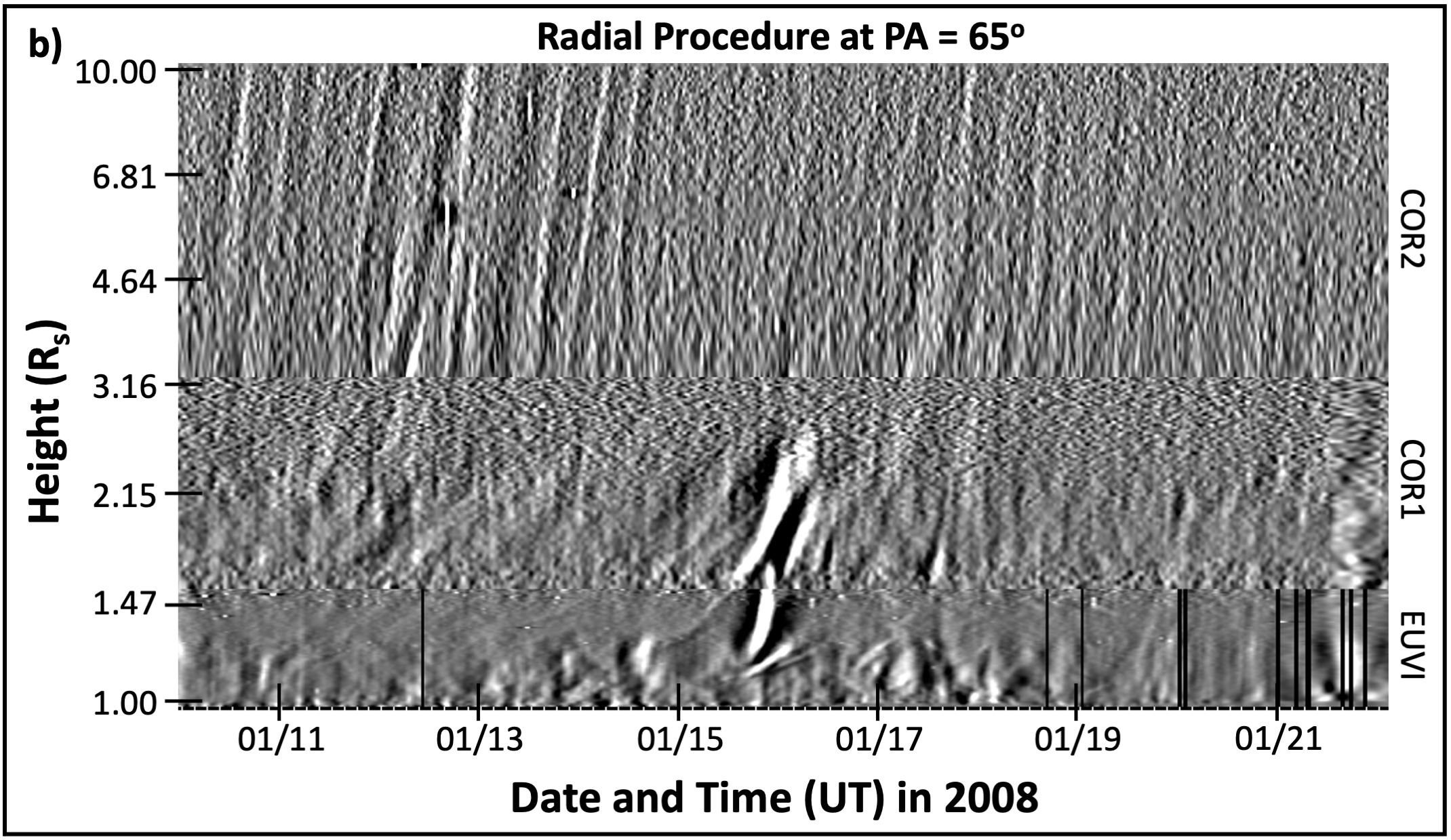}
            \includegraphics[width=0.52\textwidth]{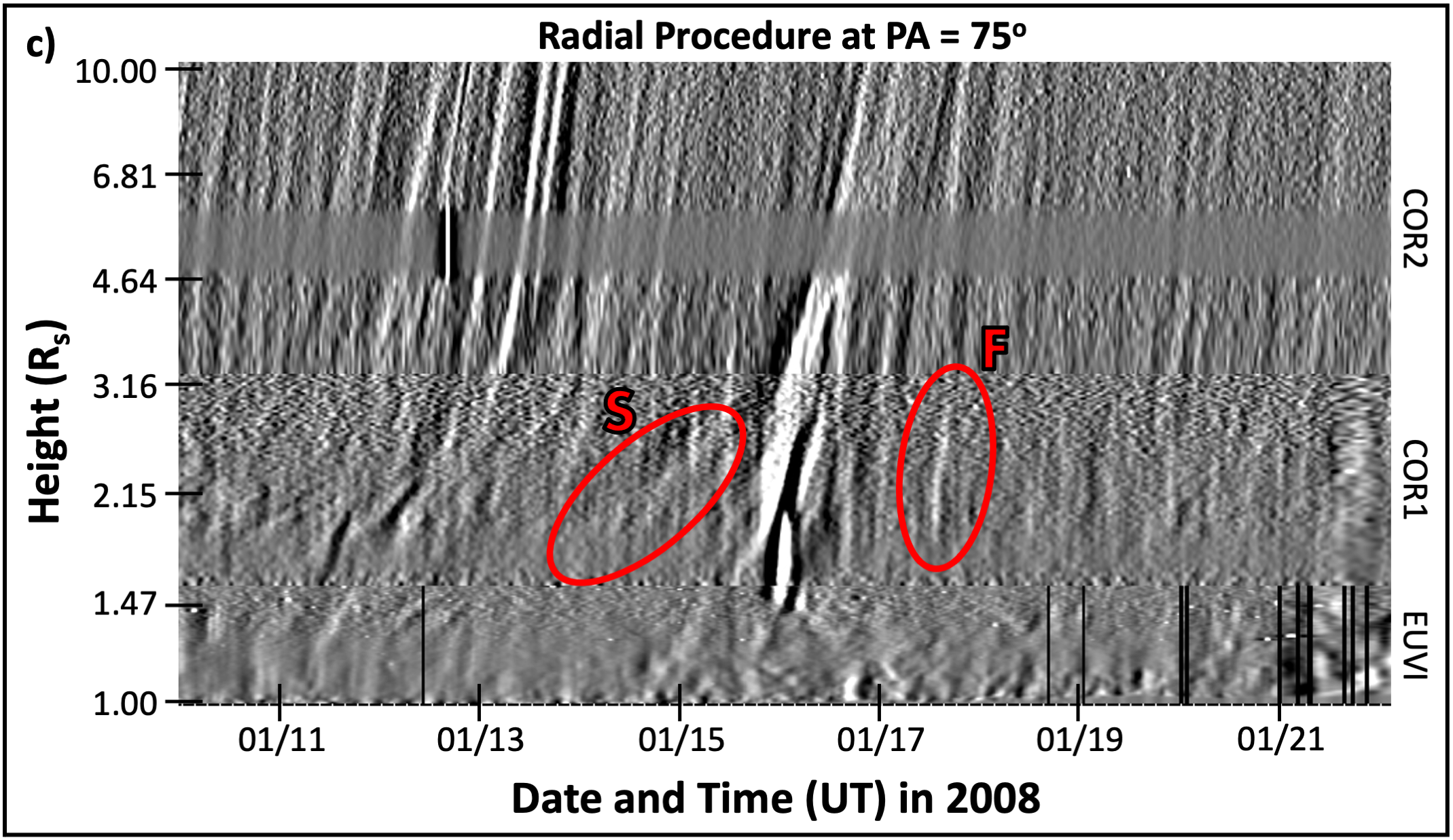}
            \includegraphics[width=0.52\textwidth]{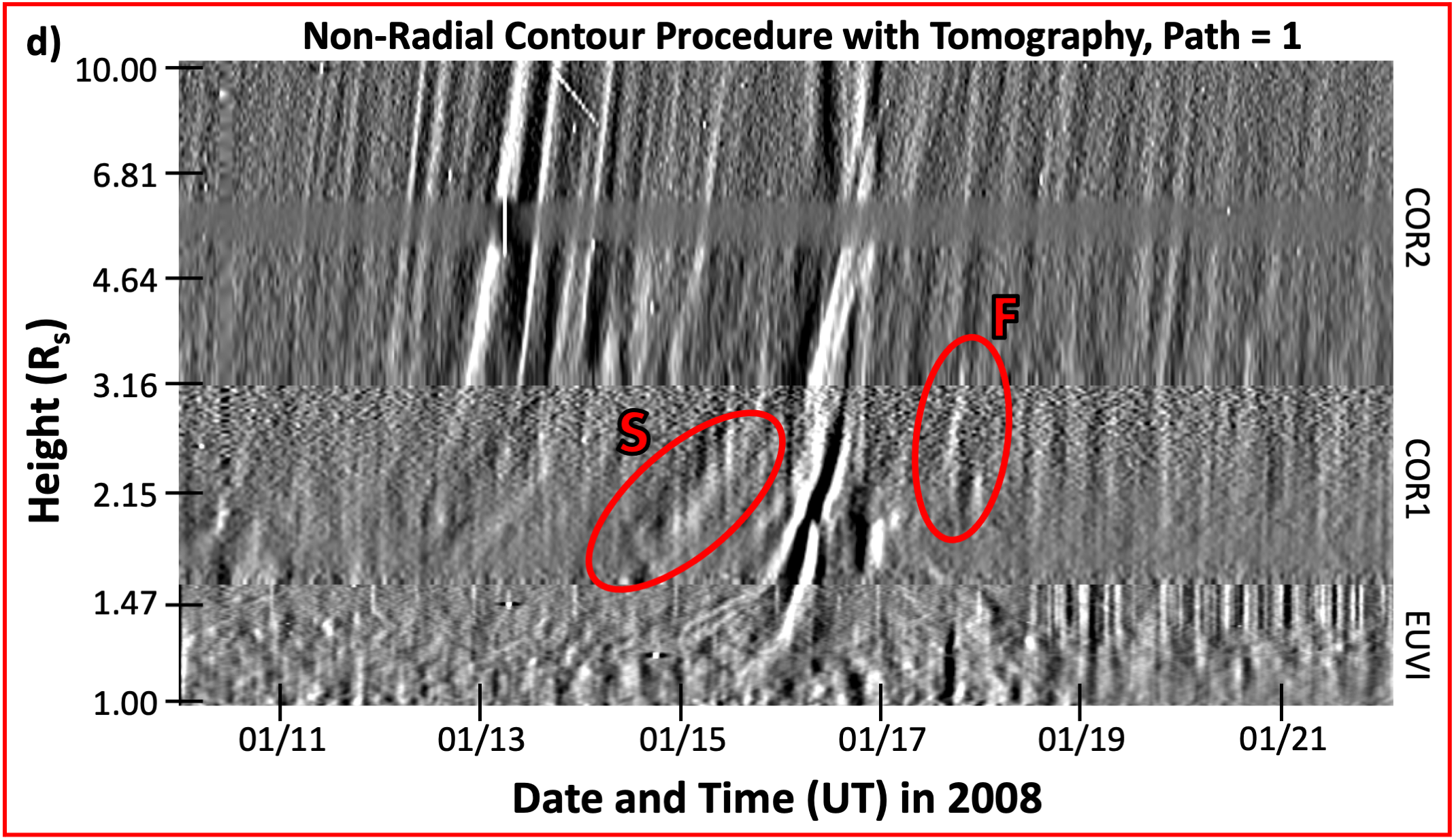}
            \caption{Comparison between \textit{radial} Ht-T plots using the method by \citet{alzate2021} and \textit{non-radial} Ht-T plots using our new method.  a-c) Radial EUVI, COR1 and COR2 composite plot for radial slices centered at a) PA=55$^o$, b) PA=65$^o$ and c) PA=75$^o$.  d) Non-radial profile using \textit{contour with tomography} method for Path=1.}
            \label{fig:rad-nonrad-htt-plots}
    \end{figure}

In Figure \ref{fig:rad-nonrad-htt-plots}, we compared the results of the non-radial Ht-T plots stemming from the \textit{contour with tomography} procedure with radial Ht-T plots centered at $pa=$55$^\circ$, 65$^\circ$, and 75$^\circ$ following the methodology of \citet{alzate2021}. We considered multiple radial slices to cover the extent of position angles from the non-radial path. Black lines in the EUVI region of the radial Ht-t plots (Panels a-c) indicate data gaps.  When generating the non-radial Ht-t plots (Panel d), the data gaps were filled by linear interpolation. The non-radial results consist of a composite of the results from the radial slices (see $pa$ values for Path 1 in Table \ref{tab:pa_ranges}).

On 15 January, the CME lift off was observed in the 55$^o$ radial slice (Panel a) as two narrow bright trajectories joining into one in the COR1 FOV. In the 65$^o$ radial slice (Panel b), a brighter portion of the CME marks its outflows on 16 January in the EUVI FOV, joining two bright trajectories in the COR1 FOV.  In the 75$^o$ radial slice (Panel c), the CME propagation is mostly confined in the COR1 and COR2 FOV. For this specific case, all these features are clearly observed in the Path 1 non-radial Ht-T plot (Panel d), which provides an uninterrupted tracking of the CME propagation.  Note that on 15-17 January, Path 1 spans $pa\sim$42.0$^\circ$–59.1$^\circ$ at $\sim$1 $R_s$ and $pa\sim$72.0$^\circ$–78.2$^\circ$ at $\sim$8 $R_s$.  For this specific CME, the non-radial methodology results in Panel d show features in EUVI starting as low as $\sim$1.1 $R_s$ and unequivocally connect to the CME in COR1, unlike the radial methodology results in Panel c.  Note that the crossing of outflows between adjacent non-radial profiles is still possible, especially for large-scale structures.  This is the case for the CME discussed above.  Signatures of the CME are also observed in Paths 2–6 (not shown) progressively starting at larger heights, possibly related to the CME expansion.

In Figure \ref{fig:rad-nonrad-htt-plots}d, we can also recognize some of the slow- and fast-moving features first discussed in \citet{alzate2021}, marked with a letter ``S" and ``F", respectively.  Even though they might appear subtle in Figure \ref{fig:rad-nonrad-htt-plots}c, they were identified also in single frame images (see Figure 5 of \citealt{alzate2021}) and, here, the non-radial Ht-T plot (Figure \ref{fig:rad-nonrad-htt-plots}d) shows the same features but brighter and more coherent.  In the hours leading up to the CME, a slow feature was observed in the radial profile with position angle of $\approx$75$^o$ at heights between 2.1 and 3.1 $R_s$ in the COR1 FOV. A similar feature is also observed in the non-radial profile along Path 1, but with a clearer trajectory down to $\approx$1.5 $R_s$.  This, again, highlights the ability of the non-radial procedure to follow outflows for a wide range of time/heights.  An example of a fast-moving feature is evident after the CME.  In the \textit{pa}$\approx$75$^o$ radial profile, we see the corresponding bright track from $\approx$2.1 $R_s$ to $\approx$3.1 $R_s$.  In the non-radial profile, the same feature appears at a larger height in the COR1 FOV ($\approx$2.2 $R_s$), but can be followed unequivocally into the COR2 FOV.


\section{Discussion} 
\label{sec:discussion}

We developed a new methodology to extract the streamer boundary profile from the solar surface up to 8.0 $R_s$ from observations by EUVI 195 \AA, COR1 and COR2.  Following the profile evolution at different observation times, we were able to investigate large spatio-temporal variations of the boundary of an equatorial streamer.  Based on our analysis and the results from each method, the \textit{contour method with tomography} works best by generating smoother profiles and non-radial Ht-T plot outflow trajectories that are clearly distinguishable across the entirety of the FOVs of interest.  Despite the few number of location points extracted from tomography reconstructions, they proved very valuable in constraining the non-radial profiles by partially addressing line-of-sight issues in observation data.  It is important to note that the criteria chosen for this analysis were tuned specifically for a period of quiet Sun in January 2008.  Analysis of other streamers and different solar corona conditions is needed to establish fixed criteria or best practice for applying this analysis.

Focusing on the results of the \textit{contour with tomography} method, we investigated the dynamics of the streamer boundaries.  Time series of the streamer boundary position angle at fixed heights manifested a trend, more evident for the South boundary, associated with the progressively wider extension in latitude of the streamer, which is recognizable in the tomography reconstruction of coronal electron density in Figure \ref{fig:obs-model-compare}f. Moving from Carrington longitude values of $\approx$127$^o$ (white filled circles around 10 January) to $\approx$330$^o$ (orange filled circles/squares around 22 January), red areas in the figure corresponding to denser regions and hence the coronal streamer, progressively expand with decreasing Carrington longitude, that is, with increasing time.

In addition to the long term trend, the time series representation of the streamer boundaries manifested fluctuations.  By applying a robust spectral analysis \citep{dimatteo2021}, we were able to reveal the periodic nature of position angle variations for portions of the streamer boundary.  We identified position angle oscillations at periods of 36–48 hours (5.8–7.7 $\mu$Hz) and 10.5–14.6 hours (19.0–26.5 $\mu$Hz), the latter at progressively lower frequencies with increasing height.  Position angle periodic variations at the higher frequency, observed mostly between 1.4-2.4 $R_s$, show larger amplitude fluctuations following the launch of the CME on January 15.  This suggests a possible relation to streamer waves, transverse motion of the streamer stalk believed to be caused by the streamer interaction with a CME \citep{chen2010,feng2011,decraemer2020}.  Streamer waves observed in white light coronagraph data have been reported at periodicities from $\lesssim$100 min \citep{chen2010,feng2011} up to 8 hours \citep{decraemer2020}.  Other large-scale disturbances related to flares and CMEs are magnetohydrodynamic (MHD) fast magnetosonic waves that, in the solar corona, can reach timescales of hours \citep{kwon2013,nakariakov2016}.  Large-scale transient coronal waves can also trigger transverse oscillations in EUV solar prominences \citep{hershaw2011,lin2011}. However, the longer period associated with kink waves in dense cool prominence threads is in the 10-100 minute range \citep{aschwanden2002,terradas2002,hershaw2011,hillier2013}, which is an order of magnitude less than the period observed in our analysis.  Consequently, the periodic streamer boundary fluctuations reported in this study are likely related to spatial deformation of the helmet streamer, even though there might be some relation of the higher frequency component with streamer waves.  

As remarked by \citet{alzate2021}, Ht-T plots generated from radial slices do not capture the non-radial nature of structures propagating between EUVI and COR1 (e.g., Figure \ref{fig:rad-nonrad-htt-plots}c misses a clear connection of the CME brightness profile to EUVI).  Structures propagating in the plane-of-sky will cross more than one radial slice and hence its trajectory will appear in plots from adjacent slices.  This results in changes in track brightness that are not indicative of the true changes in brightness of the corresponding feature.  For non-radial motion in the line-of-sight, the changes in velocity obtained from radial profiles would give the false impression that features are slowing down or speeding up.  While the latter issue still holds in this procedure, the non-radial method presented here is a new approach aimed to address plasma parcels' 2D motion in the plane-of-sky.  The non-radial Ht-T plots presented in Figures \ref{fig:all-meth-htt-plots} and \ref{fig:rad-nonrad-htt-plots}d show the full path of structures as they move from the Sun out across the FOVs of EUVI, COR1 and COR2.  Note that in some cases, it is possible to unequivocally track solar wind parcels from heliocentric distances well below 2.0 $R_s$ in an uninterrupted manner across all three FOVs.  In particular, in Figure \ref{fig:rad-nonrad-htt-plots}d we can see details of the CME from EUVI (two tracks as low as 1.1 $R_s$) to COR2 along the same non-radial path, unlike in Panel c, where only one track of the CME is clearly visible starting at 1.5 $R_s$.  This new procedure therefore provides better constraints on the identification of the source regions of specific solar wind parcels.  

The non-radial Ht-T plot confirms the occurrence of slow- and fast-moving features originally found by \citet{alzate2021}.  In Figure \ref{fig:rad-nonrad-htt-plots}d, both kinds of outflows are identified in the COR1 FOV.  Results on other non-radial paths (not shown) provide evidence of the slow- and fast-moving features also in EUVI and COR2.  One important note is that, with our new methodology, we can exclude apparent motions in the Ht-T plots due to the bright streamer rotating in the plane-of-sky \citep{alzate2021}.  This further supports the physical nature of the slow- and fast-moving features.  Further analysis of the results from other non-radial paths, focusing on small (sub-streamer size scales) features, is ongoing and will be the subject of future works.

\begin{table}[t!]
    \centering
    \begin{tabular}{ccccc}
    \hline
    \hline
    STEREO COR1 & $<v_r>$ & $v_{r,min/max}$ & $v_h$ & $v_{pa}$ \\
    \hline
    CME & $\sim$40.8  & / & 36.3$\pm$23.8  & 9.5$\pm$10.0 \\
    Slow & $\sim$4.4 & 3.2–6.4 & 10.4$\pm$5.1  & 2.1$\pm$0.7  \\
    Fast & $\sim$131 & 95–135 & 42.2$\pm$19.6  & -4.7$\pm$3.1  \\
    \hline
    \end{tabular}
    \caption{Plane-of-sky velocities, in km s$^{-1}$, estimated from COR1.  From \citet{alzate2021}, $<v_r>$ and $v_{r,min/max}$ are the average and minimum/maximum (only for slow/fast features) velocity along the \textit{radial} path. From the \textit{non-radial} Ht-T profiles, $v_h$ and $v_{pa}$ are estimates of the radial and latitudinal velocities, including standard deviation, for the CME, slow and fast features indicated in Figure \ref{fig:rad-nonrad-htt-plots}d. }
    \label{tab:Velocity_comparison}
\end{table}

Another advantage of non-radial Ht-T plots is the ability to provide estimates of outflow velocity along the radial ($v_{h}$) and latitudinal ($v_{pa}$) direction.  Here, as an example, we use the brightness profile from the propagation of the CME together with the identified slow- and fast-moving features in the COR1 FOV. We selected 11 points in the Ht-T plot along a brightness profile, assuming it is representative of the same parcel of coronal plasma, and extracted the corresponding time, height and position angle. We then computed velocities as the ratio between the change of values of the coordinates $(pa,h)$ and corresponding time for consecutive points. To reduce uncertainties, mostly related to the manual choice of the points, we repeated this approach ten times for each track and evaluated the average velocity and standard deviation.  We summarized the results in (Table \ref{tab:Velocity_comparison}) and compared them with the average value $<v_r>$ and minimum/maximum values $v_{r,min/max}$ of radial velocities for CME, slow and fast features, as estimated by \citet{alzate2021} along radial slices. We found consistency for the CME, showing comparable average velocities, and slow-features, showing $v_h$ values within the $v_{r,min/max}$ range. The $v_h$ velocity of the fast-moving feature instead is well below $<v_r>$ and outside the $v_{r,min/max}$ range reported by \citet{alzate2021}.  The present results suggest that estimates of the fast feature's velocity might be the value most affected by the non-radial plasma motion and/or the streamer evolution.  Additional insight into the nature of the CME and small-feature's propagation is provided by the latitudinal velocity.  We obtained southward propagation in the plane-of-sky for the CME and the slow-moving feature, while northward propagation for the fast-moving feature. Further statistical analysis involving the latitudinal velocity might give us new insight into the nature of these small scale features and will be the subject of future works.

\section{Conclusions} 
\label{sec:conclusions}

We presented a method for identifying streamer boundaries in EUV and WL images processed with advanced image processing techniques.  Among the combination of datasets and two boundary identification criteria (\textit{contour} and \textit{peak} methodology), the  \textit{contour with tomography} method provides the best performance in representing the streamer boundary dynamics.  Once we identified the streamer boundaries, we separated their distance at each height into seven equal parts generating non-radial paths within the streamer, along which we assume outflows move.  In this work, we showed that:

\begin{enumerate}
    \item Spectral analysis of streamer boundary position angle time series reveals the oscillatory nature of some portion of the profiles possibly related to spatial distribution of coronal electron density or the occurrence of streamer waves.
    \item The non-radial method is very effective at resolving motion of plasma parcels in the plane-of-sky near streamers.
    \item The non-radial method proves to be very valuable in providing uninterrupted tracking of outflows through the EUVI, COR1 and COR2 FOV (especially below $\leq$2.0 $R_s$). As such, this approach is extremely useful in the identification of sources of outflows and, consequently, in understanding solar wind formation imposing constraints on current solar wind models.
\end{enumerate}

In future studies, we plan to extend the analysis to other (smaller) outflows evident in the non-radial Ht-T plots.  We plan to improve the accuracy of the velocity estimates by using more robust methods and further constraining line-of-sight effects using multi-spacecraft observations.  A current effort is underway to compare the velocities of outflows in the non-radial Ht-T plots presented here to those identified by \citet{alzate2021}.

\section*{Acknowledgments}

The authors would like to thank the anonymous reviewer for their valuable and thorough comments that have improved the quality of this work.  N. Alzate acknowledges support from NASA ROSES through HGI Grant No. 80NSSC20K1070 and PSP-GI grant No. 80NSSC21K1945.  H. Morgan acknowledges STFC grant ST/S000518/1 and Leverhulme grant RPG-2019-361 to Aberystwyth University.  The work of S. Di Matteo was supported under the ECIP Grant No. 80NSSC21K0459 and PSP-GI grant No. 80NSSC21K1945. The tomography maps are available at \url{https://solarphysics.aber.ac.uk/Archives/tomography/}.


%

\vspace{5mm}
\facilities{STEREO (EUVI, COR1 and COR2). The tomography maps were produced using SuperComputing Wales.}



\software{The spectral analysis code used in this work is freely available on the Zenodo platform \citep{DiMatteo2020}}

\bibliography{Alzate_et-al}{}
\bibliographystyle{aasjournal}



\end{document}